# Nonlinear phonon Hall effects in ferroelectrics: its existence and non-volatile electrical control

W. Luo[1]†, J. Y. Ji[2]†, P. Chen[1], Yang Xu[6], L. F. Zhang[3]*, H. J. Xiang[2,4,5]* and L. Bellaiche[1]

[1]Physics Department and Institute for Nanoscience and Engineering, University of Arkansas, Fayetteville, Arkansas 72701, USA
[2]Key Laboratory of Computational Physical Sciences (Ministry of Education), State Key Laboratory of Surface Physics, and Department of Physics, Fudan University, Shanghai 200433, China
[3]Phonon Engineering Research Center of Jiangsu Province, Center for Quantum Transport and Thermal Energy Science, Institute of Physics Frontiers and Interdisciplinary Sciences, School of Physics and Technology, Nanjing Normal University, Nanjing 210023, China
[4]Collaborative Innovation Center of Advanced Microstructures, Nanjing 210093, China
[5]Shanghai Qi Zhi Institute, Shanghai 200232, China
[6]Key Laboratory of Polar Materials and Devices (MOE), School of Physics and Electronic Science, East China Normal University, Shanghai 200241, China
Corresponding author: phyzlf@njnu.edu.cn, hxiang@fudan.edu.cn
†These authors contributed equally to this work

**Abstract:** Nonlinear Hall effects have been previously investigated in non-centrosymmetric systems for electronic systems. However, they only exist in metallic systems and are not compatible with ferroelectrics since these latter are insulators, hence limiting their applications. On the other hand, ferroelectrics naturally break inversion symmetry and can induce a non-zero Berry curvature. Here, we show that a non-volatile electric-field control of heat current can be realized in ferroelectrics through the nonlinear phonon Hall effects. More precisely, based on Boltzmann equation under the relaxation-time approximation, we derive the equation for nonlinear phonon Hall effects, and further show that the behaviors of nonlinear phonon (Boson) Hall effects are very different from nonlinear Hall effects for electrons (Fermion). Our work provides a route for electric-field control of thermal Hall current in ferroelectrics.

**Introduction**

Hall effects [1-4], which typically require breaking of the time reversal (TR) symmetry, have been widely studied in the past years. Recently, the nonlinear Hall effect (NHE) [5-16], which can exist without breaking TR symmetry, attracted a lot of attention. It arises from the Berry curvature dipole (BCD) [5] in an inversion-symmetry broken system. Due to the symmetry requirement of BCD, the NHE has been proposed to detect the direction of magnetic moments [17,18], electric polarization of polar metal [19] and broadband terahertz at room temperature [20]. NHE also has been extended to nonlinear thermal Hall effects [21-25], that involve electrons and magnons.

Ferroelectrics, which naturally break inversion-symmetry, have been widely investigated due to, e.g., their applications in non-volatile memory devices [26-28]. Ferroelectrics usually are insulators. However, the response of NHE (for electronic systems) to external electric fields is a Fermi liquid property [5] and can only exist in metallic systems. Hence, ferroelectrics and NHE are naturally not compatible to each other in principles. Note that Xiao *et al*. predicted that NHE can exist in the $LiOsO_3$ ferroelectric metal [19], but it cannot be controlled by switching the polarization of metallic $LiOsO_3$ which is metallic. Similarly, although ferroelectrics can also be metallic for some special two-dimensional (2D) systems with out-of-plane polarization [29-32], the switchable ability of their polarization is also very limited, except for 2D $WTe_2$ [29]. On the other hand, compared with electronic (Fermionic) systems, phononic (Bosonic) systems may be a good platform for controlling the nonlinear phonon Hall effect (NPHE) since it is not limited to metallic systems. A naturally question thus arises, namely does the NPHE exist in ferroelectric systems? If yes, are there any differences in the temperature-dependent behavior of NHE and NPHE? Moreover, could the NPHE be switched by an external electric field? Obviously, answering these questions will not only open a new possibility for NPHE, but also may lead to thermal Hall devices integrated in ferroelectrics.

In this work, by solving the Boltzmann equation under relaxation-time approximation [33], we find that NPHE can exist in any system having broken inversion-symmetry. It is very different in nature from the NHE for electrons. For instance, the NHE for electrons are only decided by the states located near the Fermi level, while NPHE are contributed by all phonon bands. Furthermore, the nonlinear Hall conductivity for electrons increases with decreasing temperature [7,14]. In contrast, the nonlinear phonon Hall conductivity first increases with increasing temperature, and then, after reaching a maximum value at a specific temperature, decreases with further heating the system. We also find that, for three-dimensional (3D) ferroelectrics, the nonlinear phonon Hall conductivity changes its sign when the ferroelectric polarization switches. While, for 2D ferroelectrics, if the two polarization states are connected by a $M_z$ mirror symmetry, the nonlinear phonon Hall conductivity does not change its sign due to the lack of component of Berry curvature $\Omega_x, \Omega_y$, and lack of the good quantum number $k_z$ in 2D systems. We further find that the switching of ferroelectric polarization in 3D systems switches the Weyl chirality and topological charge of Weyl phonons.

**Nonlinear phonon Hall effects**

The phonon Hall effect has been previously observed in experiments [34,35]. Theoretical studies show that it is caused by the Berry curvature of phonon bands in systems with Raman spin phonon coupling [36-38]. The phonon Hall conductivity can be expressed as [39,40]

$$\kappa_{ab} = -\frac{k_B^2 T}{\hbar V} \sum_{n,q} c_2(\rho) \Omega_{n,c}(\boldsymbol{q}) \quad (1)$$

where, $\kappa_{ab}$ is the phonon Hall conductivity. $T$ and $V$ are the temperature and materials' volume. $k_B$ and $\hbar$ are Boltzmann and reduced Plank constants. $\boldsymbol{q}$ and $n$ are Bloch wave vector and band index for phonons. $a, b$ and $c$ (equivalent to $x, y, and\ z$) are Cartesian coordinates. $\rho$ and $\Omega$ are distribution function and Berry curvature for phonons. Here, $c_2(\rho)$[40] is a coefficient that depends on $\rho$, and which is different from the $c_0$ ($c_0 = \rho$) involved in NHE [5] and $c_1$ ($c_1 = (1+\rho)ln(1+\rho) - \rho ln\rho$) associated with nonlinear (spin) Nernst effect [21,24]. For non-magnetic systems (in their equilibrium state), $\kappa_{ab}$ is always zero since the integration of Berry curvature in the whole Brillouin zone (BZ) for all phonon bands vanishes. Consequently, the phonon Hall conductivity vanishes. However, for non-magnetic systems with breaking inversion-symmetry, although the integration of Berry curvature is zero in the BZ, the Berry curvature $\boldsymbol{\Omega}_n(\boldsymbol{q})$ is non-zero at a specific vector $\boldsymbol{q}$ of a band $n$. In this case, under a temperature gradient $\nabla T$ (corresponding to the non-equilibrium state), we will show that the phonon Hall conductivity $\kappa_{ab}$ can be non-zero, leading to the NPHE.

The expression of $c_2$ in Equation (1) is [40]

$$c_2(\rho) = (1+\rho)\left(\ln\frac{1+\rho}{\rho}\right)^2 - (\ln\rho)^2 - 2Li_2(-\rho) \quad (2)$$

$Li_2$ is the polylogarithm of order 2. Note that, here, we consider the non-equilibrium state which is caused by the temperature gradient $\nabla T$. Here, $\rho$ represents the distribution function for non-equilibrium state and we expand it as $\rho \approx \rho_0 + \rho_1$, where $\rho_0$ refers to the Boson-Einstein distribution function for equilibrium state, and $\rho_1$ is a small first-order term of the temperature gradient $\nabla T$.

Furthermore, the Boltzmann equation under relaxation-time approximation, can be written as [41]:

$$\frac{\partial \rho}{\partial t} = -\boldsymbol{v}_{n,\boldsymbol{q}}\frac{\partial \rho}{\partial \boldsymbol{r}}(\boldsymbol{q},\boldsymbol{r},t) - \left(\frac{d\boldsymbol{q}}{dt}\right)\frac{\partial \rho}{\partial \boldsymbol{q}}(\boldsymbol{q},\boldsymbol{r},t) - \frac{\rho-\rho_0}{\tau(\boldsymbol{q})} \quad (3)$$

where, $t$ and $\boldsymbol{r}$ represent the time and position vector, respectively. $\boldsymbol{v}_{n,\boldsymbol{q}}$ is the group velocity for band $n$ at the $\boldsymbol{q}$ point in BZ. Assuming the system in a steady state, we have $\frac{\partial \rho}{\partial t} = 0$. Since phonons have no charge, there is no electric field force. Thus, we have $\frac{d\boldsymbol{q}}{dt} = 0$. The last term of equation (3) $-\frac{\rho-\rho_0}{\tau(\boldsymbol{q})}$ represents the collision term. $\tau(\boldsymbol{q})$ represents the relaxation time for phonons, which depends on $\boldsymbol{q}$. For simplicity, we

write it from now on as $\tau$. Thus, Equation (3) can be simplified (for the $a$ component) to:

$$v_{n,a}(\boldsymbol{q})\frac{\partial \rho}{\partial r_a} = -\frac{\rho-\rho_0}{\tau} \quad (4)$$

Putting $\rho \approx \rho_0 + \rho_1$ into equation (4) and comparing the expansion coefficients in the first-order of temperature gradient, we get

$$\rho_1 = -\tau v_{n,a}(\boldsymbol{q})\frac{\partial \rho_0}{\partial T}\frac{\partial T}{\partial r_a} \quad (5)$$

Putting $\rho \approx \rho_0 + \rho_1$ into equation (2), and then using equation (5), we obtain after a straightforward but tedious calculation (see part I of SI for details):

$$c_2(\rho) = c_2(\rho_0) + \frac{\tau k_B}{\hbar}\left(\frac{E_{n,q}-\mu}{k_B T}\right)^3 \frac{\partial \rho_0}{\partial q_a}\frac{\partial T}{\partial r_a} \quad (6)$$

Where $E_{n,q}$ and $\mu$ are the energy for band $n$ at the $\boldsymbol{q}$ point in the BZ and the chemical potential, respectively. For phonons, the chemical potential $\mu$ is always zero since phonons are Bosons. Putting equation (6) into equation (1), and by using $\kappa_{ab} = -\frac{k_B^2 T}{\hbar V}\sum_{n,q} c_2(\rho_0)\Omega_{n,c}(\boldsymbol{q}) = 0$ for the equilibrium state (Note that we focus on the non-magnetic system), the temperature gradient induced thermal conductivity can be written for non-equilibrium states as

$$\kappa_{ab} = -\frac{\tau \nabla_d T}{V\hbar^2}\epsilon_{abc}\sum_{\boldsymbol{q},n}\frac{E_{n,q}^3}{T^2}\frac{\partial \rho_0}{\partial q_d}\Omega_{n,c}(\boldsymbol{q}) \quad (7)$$

Here, $\frac{\partial T}{\partial r_d}$ is denoted by $\nabla_d T$ and $d = (a,b,c)$. Since $j_a = -\kappa_{ab}\nabla_b T$, we have

$$j_a = \frac{\tau \nabla_d T \nabla_b T}{V\hbar^2}\epsilon_{abc}\sum_{\boldsymbol{q},n}\frac{E_{n,q}^3}{T^2}\frac{\partial \rho_0}{\partial q_d}\Omega_{n,c}(\boldsymbol{q}) \quad (8)$$

where $\epsilon_{abc}$ is the Levi-Civita symbol. One can see that the thermal current $j_a$ (related with phonons) is proportional to the second order of the temperature gradient. Hence, we name this effect as the NPHE.

For NHE of electrons, at the equilibrium state, $\int_q c_0(f_0)\Omega$ (In this case, $c_0 = f_0$ is the Fermi-Dirac distribution for equilibrium state) equals to zero for non-magnetic systems. Hence, by integration by parts, one gets $\int_q \frac{\partial f_0}{\partial q}\Omega = -\int_q f_0 \frac{\partial \Omega}{\partial q}$. The right side $-\int_q f_0 \frac{\partial \Omega}{\partial q}$ is the so-called BCD since it is related with the gradient of Berry curvature in the BZ. However, in our case, at the equilibrium state, we have $\int_q c_2(\rho_0)\Omega = 0$ (The form of $c_2(\rho_0)$ is shown in Eq. (2)) instead of $\int_q c_0\Omega = 0$ for non-magnetic system. Hence, we cannot use the BCD to describe the nonlinear thermal current for phonons. However, we still can define a similar pseudotensorial quantity $P_{ab} = \frac{1}{VT^2}\sum_{q,n} E_{n,q}^3 \frac{\partial \rho_0}{\partial q_a}\Omega_{n,b}(q)$ which has the same transformation properties as BCD under symmetry operations. Since $\rho_0$ is an even function under TR symmetry, the transformation properties under crystal symmetry of $P_{ab}$ is only decided by $q$ and $\Omega$ which is the same as BCD. Note that we have a factor $\frac{E_{n,q}^3}{T^2}$ for the pseudotensorial quantity $P_{ab}$, which is different from the factor $\frac{(E_{n,q}-\mu)^2}{T^2}$ for nonlinear anomalous Nernst effect [21]. Moreover, and as different from the nonlinear anomalous Nernst effect and NHE for electrons [5], the factor $\frac{\partial \rho_0}{\partial q_a}$ in $P_{ab}$ indicates that the nonlinear thermal current [Equation (8)] is contributed by all phonon bands instead of the states near the Fermi level since phonons are bosons. For 3D systems, different from the BCD which is dimensionless, the pseudotensorial quantity $P_{ab}$ has a unit of $eV * k_B^2$. While, for 2D systems, $P_{ab}$ has a unit of $eV * k_B^2 * Å$.

**Switchable NPHE in 3D ferroelectrics**

In this part, we will show that, due to the symmetry requirement of the pseudotensorial quantity $P_{ab} = \frac{1}{VT^2}\sum_{q,n} E_{n,q}^3 \frac{\partial \rho_0}{\partial q_a}\Omega_{n,b}(q)$, the NPHE must change its sign when the polarization of a 3D ferroelectric is reversed. For a 3D ferroelectric, there must exist a polar axis. For simplify, we assume it is along the $+c$ direction. Thus, the mirror symmetry $M_c$ is broken by the polarization, and one can then show that $P_{ab}$ must be non-zero. Hence, one can see that the ferroelectric polarization is highly related with the symmetry of the pseudotensorial quantity $P_{ab}$. When the polarization is switched, the structure is reflected by inversion symmetry. Under inversion symmetry, $\mathbf{\Omega}_b$ is even and $\mathbf{q}_a$ is odd (Note that $\rho_0$ is even under TR symmetry). Thus, the $P_{ab}$ must reverse its sign when the polarization is reverted. More interestingly and since the Berry curvature does not change sign under inversion symmetry, the Weyl chirality and topological charge of Weyl points for 3D ferroelectrics should reverse their sign when switching the electrical polarization of 3D ferroelectrics. Related details of such features can be found in part II of SI.

We then take the 3D ferroelectric PbTiO$_3$[42] as an example and explain how the NHPE can be controlled by switching the polarization. The crystal structure of PbTiO$_3$ is shown in Fig. 1(a). It is a tetragonal phase with the $P4mm$ space group (number 99) and $C_{4v}$ point group. This latter point group implies that the pseudotensorial quantity $P_{ab}$ has the form (see detailed analysis in part III of SI) of

$$\begin{bmatrix} 0 & P_{xy} & 0 \\ -P_{xy} & 0 & 0 \\ 0 & 0 & 0 \end{bmatrix}$$

For which only $P_{xy}$ and $P_{yx}$ are non-zero. Based on equation (8), we have

$$j_z = \frac{\tau(\nabla_x T)^2}{V\hbar^2} \sum_{q,n} \frac{E_{n,q}^3}{T^2} \frac{\partial \rho_0}{\partial q_x} \Omega_{n,y}(\boldsymbol{q}) = \frac{\tau(\nabla_x T)^2}{\hbar^2} P_{xy} \quad (9)$$

This equation implies that, for PbTiO$_3$ with a polarization down, adding a temperature gradient ($\nabla_x T$) along the $-x$ direction will induce a thermal current (along the $-z$ direction) that is proportional to both $(\nabla_x T)^2$ and $P_{xy}$ ($P_{xy}$ is positive for a polarization down with temperature higher than 30 Kelvin (K)) [see Fig. 1(a)]. After the polarization is switched by an electric field, the nonlinear thermal current $j_z$ switches its sign due to the change of sign of $P_{xy}$, while the temperature gradient is maintained.

To compute the pseudotensorial quantity $P_{ab} = \frac{1}{VT^2}\sum_{q,n} E_{n,q}^3 \frac{\partial \rho_0}{\partial q_a} \Omega_{n,b}(\boldsymbol{q})$ for PbTiO$_3$, we needed to calculate the phonon Berry curvature $\Omega_{n,b}(\boldsymbol{q})$ and derived the Bose-Einstein distribution ($\frac{\partial \rho_0}{\partial q_a}$) for the equilibrium state. Both depend on the phonon spectrum, which is shown in Fig. S8 of SI for two opposite polarization states of PbTiO$_3$. There is no imaginary part, indicating that this compound is dynamically stable. Fig. 1(c) shows the component $P_{xy}$ as a function of temperature for PbTiO$_3$ (note that (i) the temperature appears in the expression indicated above for $P_{ab}$ via $\frac{1}{T^2}$ but also via $\frac{\partial \rho_0}{\partial q_a}$; and (ii) the Curie temperature for PbTiO$_3$ is about 768K [43], which is therefore the temperature below which the nonlinear Hall current can be switched in PbTiO$_3$). One can see that its behavior is very different from the BCD as a function of temperature for electronic systems [14]. As a matter of fact, the BCD can be written as $D_{ab} = \frac{1}{V}\sum_{k,n} \frac{\partial f_0}{\partial k_a} \Omega_{n,b}(\boldsymbol{k})$ for electronic systems with $\frac{\partial f_0}{\partial k_a} = -\frac{1}{e^{\left(\frac{E_{n,k}-\mu}{k_B T}\right)} + e^{-\left(\frac{E_{n,k}-\mu}{k_B T}\right)}+2} * \frac{1}{k_B T}$.

Hence, the behavior of BCD change with temperature is decided by $-\frac{1}{e^{\left(\frac{E_{n,k}-\mu}{k_B T}\right)} + e^{-\left(\frac{E_{n,k}-\mu}{k_B T}\right)}+2} * \frac{1}{k_B T}$. Consequently, when increasing the temperature, the BCD decreases monotonically for electronic system. In contrast, for phonon systems, $P_{ab} = \frac{1}{T^2}\sum_{q,n} E_{n,q}^3 \frac{\partial \rho_0}{\partial q_a} \Omega_{n,b}(\boldsymbol{q})$ with $\frac{\partial \rho_0}{\partial q_a} = -\frac{1}{e^{\left(\frac{E_{n,q}}{k_B T}\right)} + e^{-\left(\frac{E_{n,q}}{k_B T}\right)}-2} * \frac{1}{k_B T}$. Hence, the behavior of $P_{ab}$ change with temperature is decided by $-\frac{1}{e^{\left(\frac{E_{n,q}}{k_B T}\right)} + e^{-\left(\frac{E_{n,q}}{k_B T}\right)}-2} * \frac{1}{k_B T} * \frac{E_{n,q}^3}{T^2}$.

As a result, when the temperature is low (that is, lower than 20K, see Fig. 1(c)), the factor $\frac{1}{e^{\left(\frac{E_{n,q}}{k_B T}\right)} + e^{-\left(\frac{E_{n,q}}{k_B T}\right)}-2}$ is dominating and small. Hence, the magnitude of $P_{xy}$ is small as well. With increasing temperature, the factor $\frac{1}{e^{\left(\frac{E_{n,q}}{k_B T}\right)} + e^{-\left(\frac{E_{n,q}}{k_B T}\right)}-2}$ then increases faster than $\frac{1}{k_B T} * \frac{E_{n,q}^3}{T^2}$ decreases, and, therefore, the magnitude of $P_{xy}$ quickly increases (between 20K and 80K). In the high temperature limit, the factor $\frac{1}{e^{\left(\frac{E_{n,q}}{k_B T}\right)} + e^{-\left(\frac{E_{n,q}}{k_B T}\right)}-2}$ increases slower than $\frac{1}{k_B T} * \frac{E_{n,q}^3}{T^2}$ decreases, resulting in the magnitude of $P_{xy}$ decreasing. One can thus see that the different behavior between BCD $D_{ab}$ and $P_{ab}$ is due to their different distribution function, and the fact that the $P_{ab}$ has one more factor of $\frac{E_{n,q}^3}{T^2}$ compared with $D_{ab}$[5]. Fig. 1(c) also reveals that there is a specific temperature at which the $P_{xy}$ reaches the maximum value, which can be a feature to look for when searching experimental evidence of NPHE (Note that, different materials should have different "*specific temperature*"). In addition, the pseudotensorial quantity $P_{ab}$ changes its sign as the temperature changes which is distinguished from the BCD behaviors[14]. Moreover, as also indicated in Fig. 1(c), reverting the polarization direction results in the $P_{xy}$ being reversed too.

We also investigated the component $P_{xy}$ as a function of frequency (the chosen range of frequency for PbTiO$_3$ is from $0\ THz$ to $25\ THz$ since the maximum frequency is around $24\ THz$), see Fig. S8 of SI). The result is shown in Fig. 1(d). The temperature is fixed at either $k_BT = 2.6\ meV$ (which corresponds to 30 K) or $k_BT = 18.1\ meV$ (corresponding to 210 K). Comparing the result for these two different temperatures, one can see that, at low temperature, only low-frequency phonon bands make contribution for $P_{xy}$; the high-frequency phonons being thus frozen. On the other hand, for higher temperature, the high-frequency phonons are excited and have significant contribution for $P_{xy}$, implying that is contributed by all phonon bands. Due to the cancellation of the Berry curvature for different phonon bands, $P_{ab}$ can change its sign for different frequency values. Note also that $P_{ab}$ at frequency $w$ is calculated by using $P_{ab} = \frac{1}{VT^2} \int_0^w \sum_{q,n} E_{n,q}^3 \frac{\partial \rho_0}{\partial q_a} \Omega_{n,b}(q))$ and summing over the states between $0\ THz$ and $w\ THz$ ($w$ is the frequency of phonons). To make sure that the computed Berry curvature is converged, we test the results for different $q$ grids (see method part in SI). The results are shown in Fig. S9 of SI.

**NPHE in 2D ferroelectrics**

For 2D systems, the Berry curvature just has a finite component for $\Omega_z$. The good quantum number for $\boldsymbol{q}$ are $q_x$ and $q_y$. Hence, only $P_{xz}$ and $P_{yz}$ can be defined. For this reason, the switching of the electrical polarization of a 2D ferroelectric will not always changes the sign of the NPHE. Specifically, a 2D ferroelectric only maintaining an out-of-plane polarization (e.g., along $+z$) cannot realize the sign reversal of the nonlinear thermal current. In fact, for a 2D ferroelectric only maintaining an out-of-plane polarization is such as its two polarization states are connected by a mirror symmetry $M_z$, with $\Omega_z, q_x$ and $q_y$ all being even under the $M_z$ operation. Thus, $P_{xz}$ and $P_{yz}$ do not change their sign when changing the polarization. One specific example is the monolayer $CuInP_2S_6$ for which the two ferroelectric states (with only out-of-plane polarization) are connected by $M_z$[44,45]. On the other hand, a 2D ferroelectric only maintaining an in-plane polarization [e.g., along $+y$ ($+x$)] can realize the sign reversal of the NPHE since its two polarization states are connected by a mirror symmetry $M_y$ ($M_x$). $\Omega_z$ being odd under the $M_y$ ($M_x$) operation while $q_x$ ($q_y$) is even under the $M_y$ ($M_x$) operation, $P_{xz}$ ($P_{yz}$) changes its sign when reverting the polarization. One example is 2D ferroelectric SnTe [46] that only has an in-plane polarization (the phonon spectra are shown in Fig. S10 of SI ). Note that the point group of monolayer SnTe is $C_{2v}$, for which there is a mirror symmetry $M_x$ whose normal is perpendicular to the $x$ axis (See Fig.2 (a)). Due to such mirror symmetry $M_x$, $P_{yz}$ vanishes and only $P_{xz}$ is nonzero. This nonzero $P_{xz}$ implies that, if there exist a temperature gradient along -$x$ ($\nabla_x T$) direction, a nonlinear thermal current will emerge along the $+y$ direction (since $P_{xy}$ is negative for a polarization down ($P-$) with temperature higher than 40 K [see Fig. 2(c)]) which is proportional to $(\nabla_x T)^2$ (see Fig. 2(a)). Switching the polarization will reverse the direction of nonlinear thermal current $j_y$ (see Fig. 2(b)). The $P_{xz}$ (for both $P+$ and $P-$ up and down polarization states, respectively) as a function of temperature and frequency are shown in Fig. 2(c) and Fig. 2(d). Note that these results originate from the use of $P_{ab} = \frac{1}{VT^2}\int_0^w \sum_{q,n} E_{n,q}^3 \frac{\partial \rho_0}{\partial q_a}\Omega_{n,b}(q)$ and that the ferroelectric transition temperature for monolayer SnTe is 275 K [46], while the range of frequency for SnTe is from $0THz$ to $5THz$ (see Fig. S10 of SI). For the 2D ferroelectrics with both in-plane and out-of-plane polarization, the nonlinear phonon Hall current can also be reverted since they have non-zero in-plane component polarization. One particular example is the 2D ferroelectric $In_2Se_3$[47].

**Conclusions**

In this work, we investigated the nonlinear phonon Hall effects based on the Boltzmann equation under relaxation-time approximation and compared differences between NHE and NPHE. Due to the looser requirement (nonlinear phonon Hall effects can occur in insulators), the nonlinear phonon Hall current and the Weyl chirality of phonons (topological charge) can be reversed by switching the ferroelectric polarization. We hope that these predictions will be observed by experiments soon and will be used to design novel quantum phononics devices.

**Acknowledgement**

The work is supported by the Vannevar Bush Faculty Fellowship (VBFF) grant no. N00014-20-1-2834 from the Department of Defense. Yang Xu is supported by the NSFC (12274125) and the Shanghai Pujiang Program (21PJ1403100). Lifa Zhang is supported by NSFC (11890703, 11975125). J. Y. Ji and H. J. Xiang are supported by NSFC (11825403, 12188101). These simulations were done using the Arkansas High Performance Computing Center. We thank Dr. L.Y.Gao for helpful discussions.

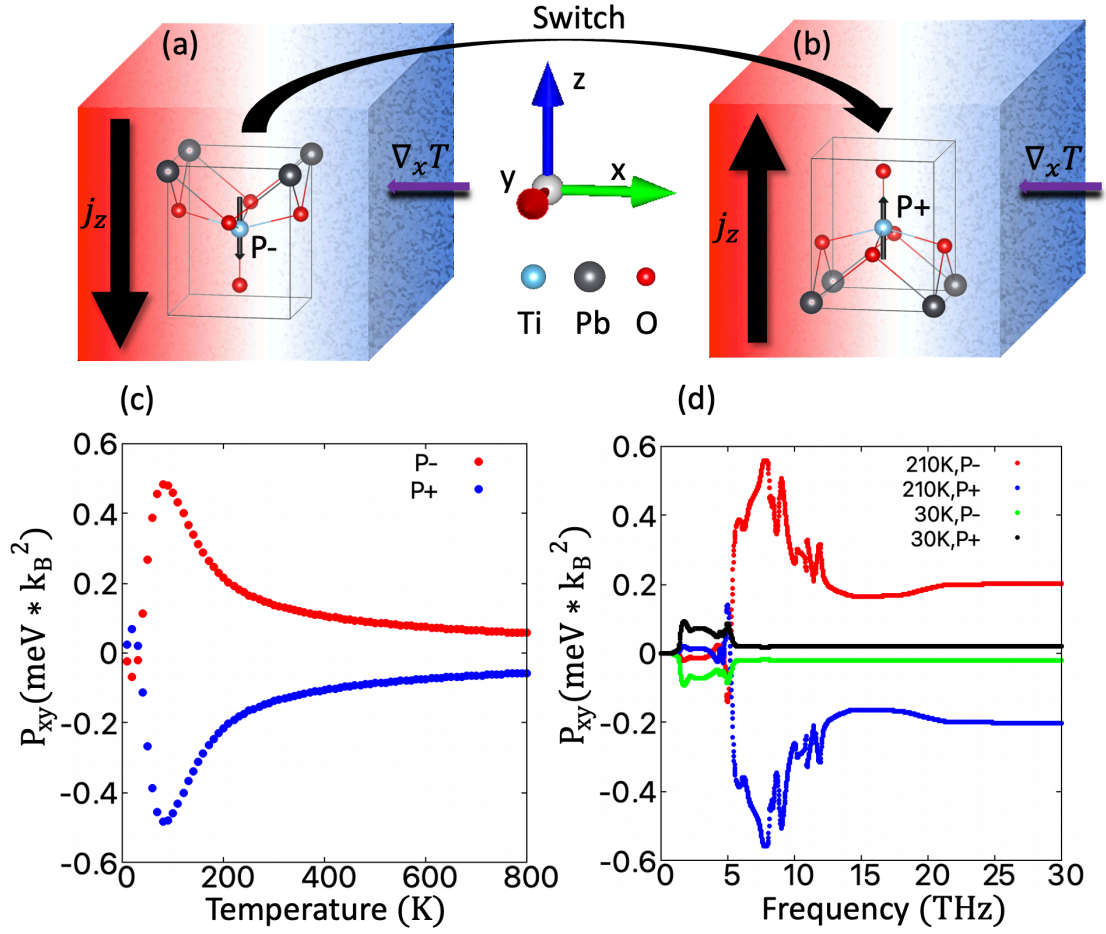

FIG. 1. Schematic diagram for the electric field control nonlinear thermal current in PbTiO$_3$ with polarization down state (a) and polarization up state (b). Red and blue region indicate the hot and cold region. $P-$ and $P+$ are used to label the polarization down and polarization up state. (c) $P_{xy}$ as a function of temperature. (d) $P_{xy}$ as a function of frequency.

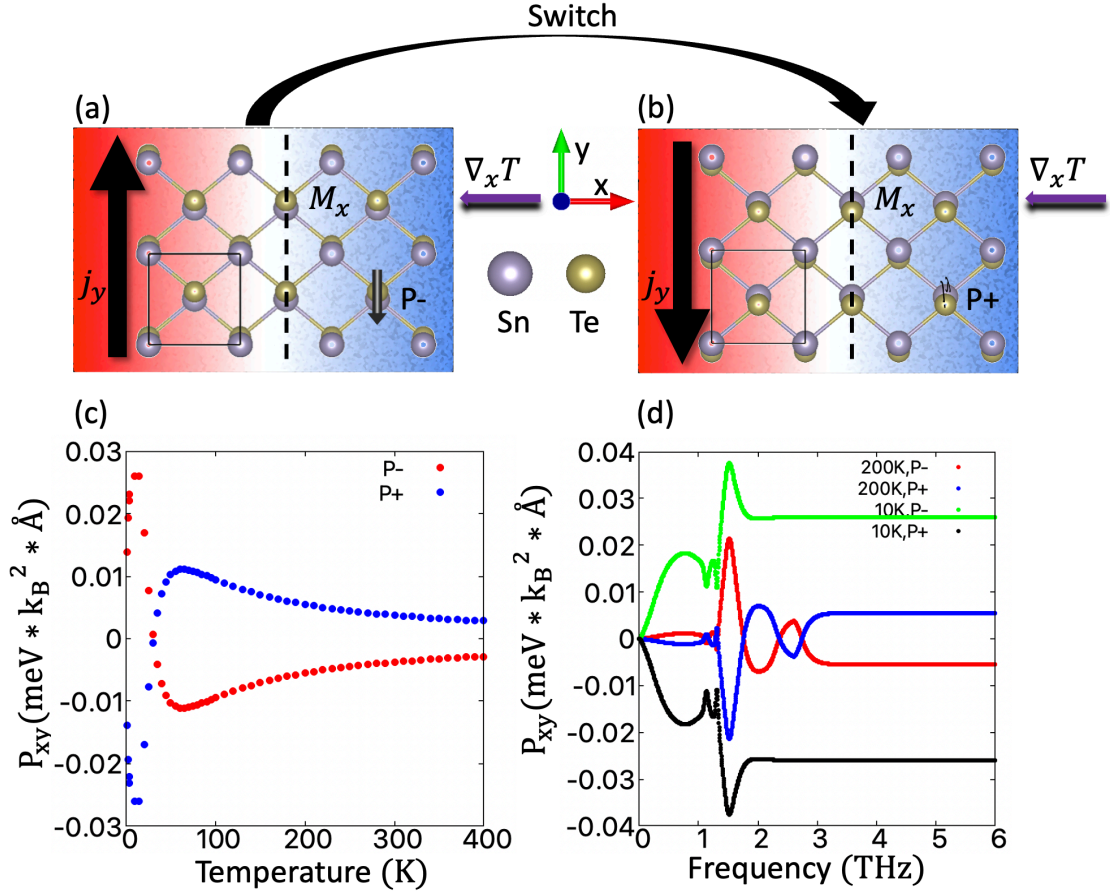

FIG.2. Schematic diagram for the electric field control nonlinear thermal current in 2D SnTe with polarization down state (a) and polarization up state (b). The black solid square line indicates the unit cell. The black dashed line represents the $M_x$ mirror symmetry. Red and blue region represent the hot and cold regions. The $P-$ and $P+$ states are connected by the $M_y$ mirror symmetry. (c) $P_{xz}$ as a function of temperature. (d) $P_{xz}$ as a function of frequency.